\documentclass[twoside,twocolumn,10pt,prb]{revtex4} 
\usepackage[T1]{fontenc}
\usepackage[utf8]{inputenc}
\usepackage{textcomp}
\usepackage{amsmath}
\usepackage{graphicx,xcolor}
\usepackage{overpic}
\usepackage[english]{babel}

\begin{document}

\title{Enhanced optical Kerr effect method for a detailed characterization of the third order nonlinearity of 2D materials applied to graphene}
\author{Evdokia Dremetsika}
\affiliation{Universit\' e libre de Bruxelles, OPERA-Photonics Group, 50 Avenue F.D.Roosevelt CP 194/5 1050 Bruxelles Belgium}
\author{Pascal Kockaert}
\email{Pascal.Kockaert@ulb.ac.be}
\affiliation{Universit\' e libre de Bruxelles, OPERA-Photonics Group, 50 Avenue F.D.Roosevelt CP 194/5 1050 Bruxelles Belgium}

\begin{abstract}
Using an enhanced optically heterodyned optical Kerr effect method and a theoretical description of the interactions between an optical beam, a single layer of graphene, and its substrate, we provide experimental answers to questions raised by theoretical models of graphene third-order nonlinear optical response. In particular, we measure separately the time response of the two main tensor components of the nonlinear susceptibility, we validate the assumption that the out-of plane tensor components are small, and we quantify the optical impact of the substrate on the measured coefficients. Our method can be applied to other 2D materials, as it relies mainly on the small ratio between the thickness and the wavelength.
\end{abstract}

\maketitle

\section{Introduction}

During the last decade, extensive research has been performed on graphene and other 2D materials for applications in photonics and optoelectronics~\cite{yu_2d_2017}. The third-order nonlinear optical response of graphene has been investigated by many groups, theoretically~\cite
{mikhailov_nonlinear_2008,%
 ooi_waveguide_2014,%
 cheng_third_2014,%
 cheng_third-order_2015,%
 semnani_nonlinear_2016,%
 mikhailov_quantum_2016,%
 mikhailov_nonperturbative_2017},
and experimentally~\cite
{hendry_coherent_2010,%
 hong_optical_2013,%
 zhang_z-scan_2012,%
 chen_nonlinear_2013,%
 demetriou_nonlinear_2016,%
 dremetsika_measuring_2016,%
 alexander_electrically_2017}. 
More recently, other 2D materials have also been studied~\cite{woodward_characterization_2017,torres-torres_third_2016}. 
Regarding the research-trend combining graphene or 2D-heterostructures with integrated photonics~\cite{lin_2d_2016,yu_2d_2017}, characterization of 2D-nonlinearities will play a key role in current and future progress in photonics. 

The tensor nature of the nonlinear susceptibility of graphene has not yet been studied, although it is a parameter that could influence the observed nonlinearity, for example, in waveguiding structures that can involve in- and out-of-plane components.

In this paper we address the theoretical hypothesis~\cite
{mikhailov_nonlinear_2008,%
 stauber_optical_2008,%
 ooi_waveguide_2014,%
 cheng_third_2014,%
 cheng_third-order_2015,%
 semnani_nonlinear_2016,%
 mikhailov_quantum_2016,%
 mikhailov_nonperturbative_2017} 
that the nonlinear optical response of graphene is limited to in-plane components. This apparently simple question is still open from an experimental point of view. It is indeed a long journey to provide an experimental answer by a direct measurement performed on a single layer sample. A first reason is that a proper modeling of the interaction of an electromagnetic wave with a 2D material reveals that textbook expressions are incomplete, and that an extended theory should be used~\cite{sipe_new_1987,felderhof_electromagnetic_1987}. Secondly, as existing methods to probe 2D materials do not provide access to the real and imaginary parts of the third-order susceptibility tensor, it is also necessary to define a new experimental scheme.

Up to now, the anisotropy induced by a pump beam on a graphene sample has been studied for different polarization angles between the pump and the probe~\cite{mittendorff_anisotropy_2014,yan_evolution_2014}, but there is no experimental work 
studying separately the two main tensor components used in the theory. Multiple reasons can probably explain this, including the very low signal provided by mono or few-layer graphene samples, the limited possibilities to probe various tensor components with Z-scan, which is a single-beam method, and the fact that only the magnitude of the fast component of \(\chi^{(3)}\) is accessible in four-wave mixing experiments. 

As explained in Ref.~\onlinecite{dremetsika_measuring_2016}, the optically heterodyned optical Kerr effect method (OHD-OKE)~\cite{smith_optically-heterodyne-detected_2002,dremetsika_measuring_2016} has many advantages over other widely used methods. In this paper, we implement an enhanced version called 2D-OHD-OKE, in which the sample is tilted, and linear polarization angles are tunable; and we provide the theoretical framework to extract tensor components from phase and amplitude jumps. 
\begin{figure}[h]\centering
\begin{overpic}[width=0.49\linewidth]{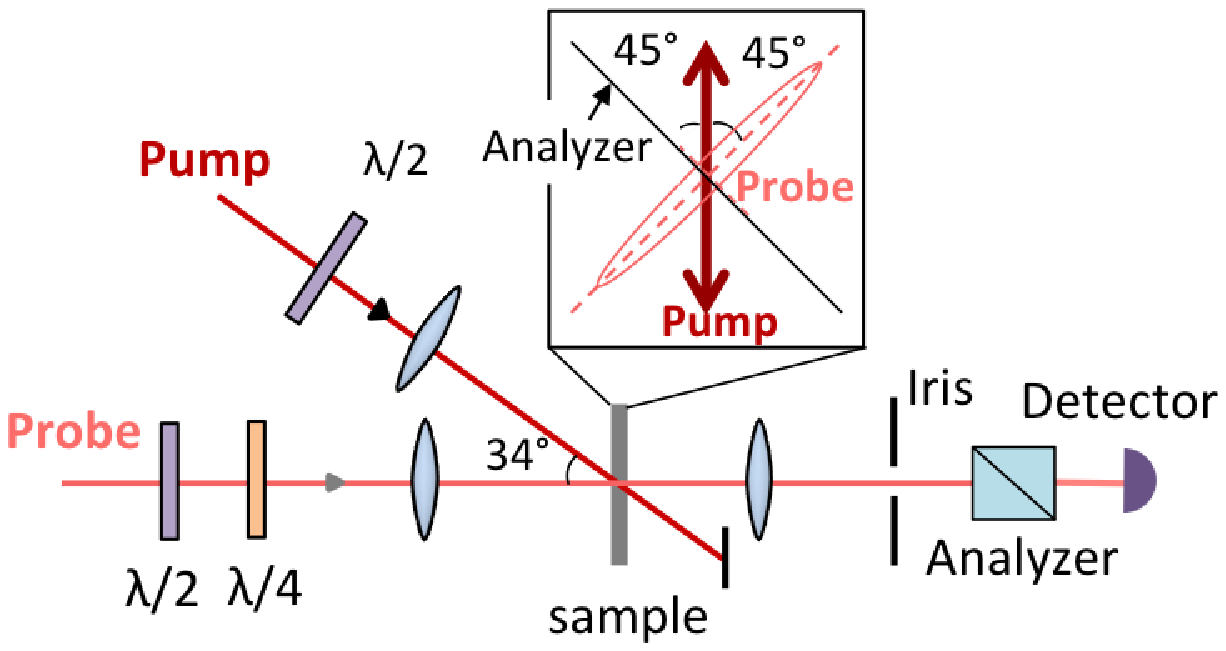}
\put(15,50){(a)}
\end{overpic}
\hfill
\begin{overpic}[width=0.49\linewidth]{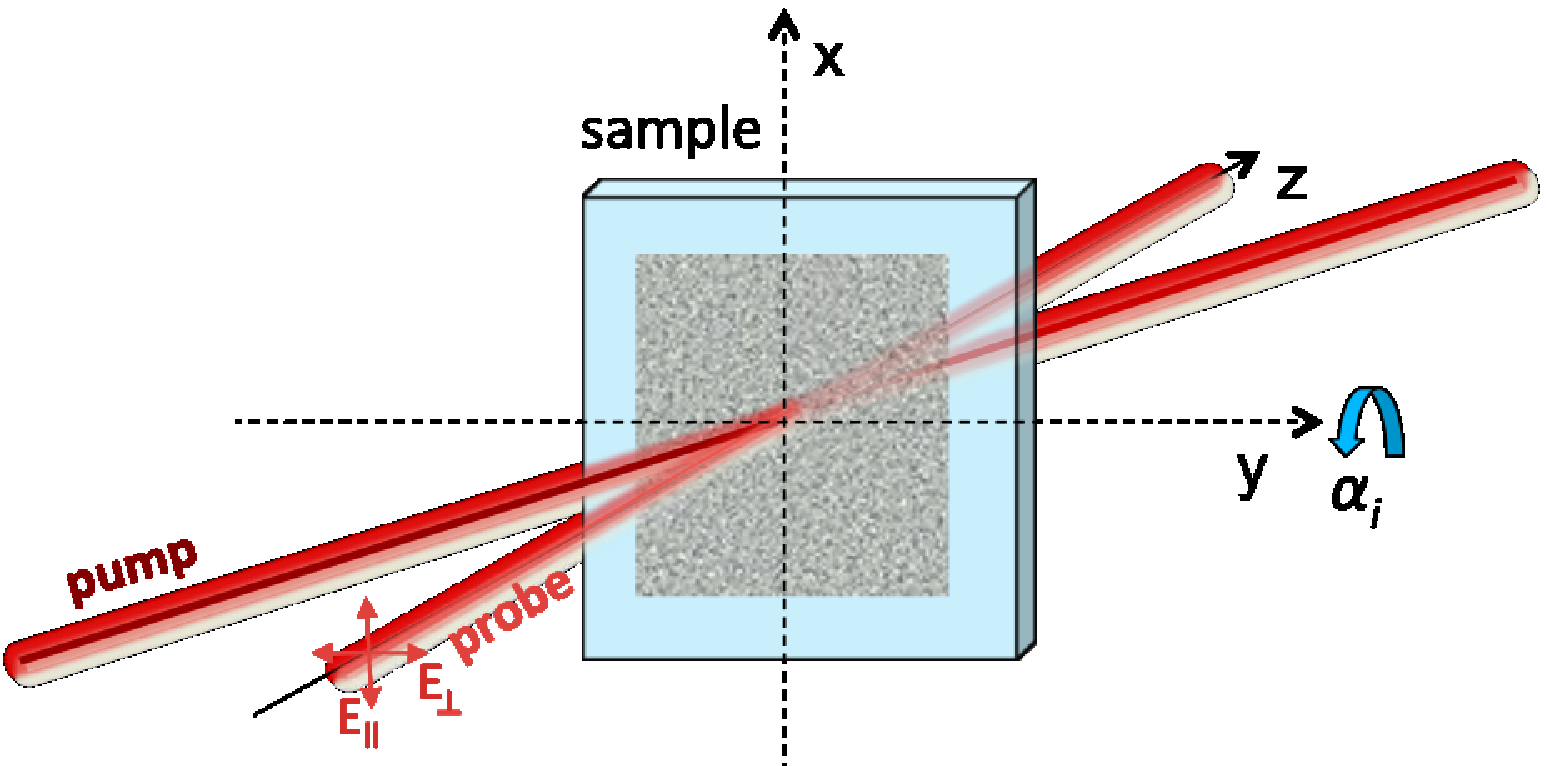}
\put(15,50){(b)}
\end{overpic}
\caption[2D-OHD-OKE setup]
        {(a)~Interaction of the pump and probe beams with the sample. Half- and quarter-waveplates are denoted respectively by \(\lambda/2\) and \(\lambda/4\). Iris denotes an iris diaphragm. (b)~The pump and probe beams define the horizontal plane that intersects the sample along the \(y\)-axis. Sample tilt around \(\hat{y}\) is denoted by \(\alpha_i\).
      }
\label{Fig:setup}
\end{figure} 

In the method section, we model the interactions of a plane wave with a graphene sheet and provide the expression of the 2D-OHD-OKE signal, taking into account new configurations that provide access to the real and/or imaginary parts of the tensor components of the nonlinear susceptibility. In the experimental section, we provide the temporal response of the tensor susceptibilities that can be accessed 
with our setup 
on a monolayer graphene on glass sample. In this way, we achieve our goal, and demonstrate that the measured out-of-plane components of the third-order nonlinear susceptibility tensor are negligible. We discuss the results and conclude on the use of 2D-OHD-OKE to access nonlinear optical parameters of any 2D material.

\section{Methods}

In OHD-OKE, a pump and a probe pulse are focused on the sample (See~\onlinecite{dremetsika_measuring_2016}, Fig.~1). The third order nonlinearity of the sample is recorded by measuring the polarization changes induced by the pump on the weak probe pulse. As this polarization change is very small, optical heterodyning using a weak phase-shifted part of the probe is performed. Fig.~\ref{Fig:setup} shows the interaction of the two beams with the sample. We calculate the phase and amplitude changes by means of boundary conditions integrating the linear and nonlinear polarization response of the graphene surface.

\subsection{Boundary conditions}
\label{Sec:BoundaryConditions}
Starting from the modeling of the sheet currents~\cite{sipe_new_1987} at the planar interface between two semi-infinite media \((a)\) and \((b)\), we write each field \(\vec{F}\) as
\begin{equation}
 \vec{F}=\vec{F}_a(x,y,z) H(-z) + \vec{\mathcal{F}}(x,y) \delta(z) + \vec{F}_b(x,y,z) H(z),
\label{Eq:Ffield}
\end{equation}
with \(H\) the Heaviside step function and \(\delta\) the Dirac distribution, where the surface field \(\vec{\mathcal{F}}\) is located at \(z=0\) and varies in the \((x,y)\) graphene plane. Inserting fields with these expressions in Maxwells' equations and collecting terms multiplying the same functions \(H(z)\), \(H(-z)\), \(\delta(z)\) and \(\delta^\prime(z)\)  provides  elegantly the plane wave solutions in the half-spaces defined by \(z<0\) and \(z>0\), as well as the boundary conditions at the interface.
Assuming that media \((a)\) and \((b)\) are dielectric and isotropic, they are characterized by real refractive indices \(n_a\) and \(n_b\). 
Writing the electric-field \(\vec{E}\), the displacement field \(\vec{D}\), the polarization field \(\vec{P}\), and the magnetic field \(\vec{B}\) as in~(\ref{Eq:Ffield}), we get the boundary conditions 
\begin{eqnarray}
 \vec{B}_b-\vec{B}_a &=& -\hat{z}\times\mu_0\partial_t\vec{\mathcal{P}}
,\label{Eq:BC_B}\\
 \hat{z}\times\left(\vec{E}_b-\vec{E}_a\right) &=& -\left(\hat{z}\times\vec{\nabla}\right)\mathcal{P}_z
,\label{Eq:BC_E}\\
 \hat{z}\cdot\left(\vec{D}_b-\vec{D}_a\right) &=& -\vec{\nabla}\cdot\vec{\mathcal{P}}
,\label{Eq:BC_D}
\end{eqnarray}
which differ from those in Ref.~\onlinecite{stauber_optical_2008}, in that the polarization induced in the graphene sheet can have components along the \(z\)-axis. This accounts for the extent of the orbitals on both sides of the graphene sheet. Starting from (\ref{Eq:BC_B})-(\ref{Eq:BC_D}), and following the classical derivation of Fresnel coefficients (see Sec.\,1.5 of Ref.~\onlinecite{born1999principles}) we get the transfer matrix
\begin{eqnarray}
E^\parallel_t
&{=}&2\left(\textstyle\frac{n_b}{n_a}+\frac{\cos\alpha_t}{\cos\alpha_i}\right)^{-1}
   \left[\textstyle
   E^\parallel_i
 + i\frac{k_0}{2n_a}\frac{\mathcal{P}_x}{\varepsilon_0}
 - i\frac{k_0n_a}{2}\tan\alpha_i\frac{\mathcal{P}_z}{\varepsilon_0}\right]
,\nonumber\\
\label{Eq:transfer_par}
\\
E^\perp_t
&{=}&2\left(\textstyle1+\frac{n_b}{n_a}\frac{\cos\alpha_t}{\cos\alpha_i}\right)^{-1}
   \left[\textstyle
   E^\perp_i
 + i\frac{k_0}{2n_a\cos\alpha_i}\frac{\mathcal{P}_y}{\varepsilon_0}\right],
\label{Eq:transfer_per}
\end{eqnarray}
where \(\parallel\) and \(\perp\) denote respectively the components parallel and orthogonal to the incidence plane of the incident and transmitted fields \(E_i\), and \(E_t\), and \(\alpha_i\) (resp. \(\alpha_t\)) is the angle between \(\hat{z}\) and the incident (resp. transmitted) wave vector. They verify the Snell-Descartes relation \(n_a\sin\alpha_i=n_b\sin\alpha_t\). For later use, we define the coefficients \(M,N,P,Q,R\) so that
\begin{equation}
 \begin{bmatrix}
  E^\parallel_t\\
  E^\perp_t
 \end{bmatrix}
=\begin{bmatrix}
  P&0\\
  0&M
 \end{bmatrix}
 \begin{bmatrix}
  E^\parallel_i\\
  E^\perp_i  
 \end{bmatrix}
+\frac{1}{\varepsilon_0}
 \begin{bmatrix}
  Q&0&R\\
  0&N&0
 \end{bmatrix}
 \begin{bmatrix}
 \mathcal{P}_x\\\mathcal{P}_y\\\mathcal{P}_z
 \end{bmatrix}
,\label{Eq:TransMNPQR}
\end{equation}
where the coefficients \(Q\), \(N\)  and \(R\) are in the order of \(k_0\).

\subsection{Linear material response}
\label{Sec:LinearMaterialResponse}

The main difference between a model of graphene using bulk or surface parameters appears in the constitutive relations. Indeed, in the bulk approach, the polarization of the medium is calculated with respect to the field transmitted in the graphene, using (\ref{Eq:transfer_par}), and~(\ref{Eq:transfer_per}) with \(n_b=n_g\) the refractive index of graphene and \(\mathcal{P}=0\). In the surface approach, we should consider the total electric field surrounding the surface sheet, as explained in Refs.~\onlinecite{felderhof_electromagnetic_1987}, and~\onlinecite{TASSIN20124062}. In what follows, we calculate this field as the symmetric combination~\cite{felderhof_electromagnetic_1987} of the incident (\(\vec{E}_i\)), reflected (\(\vec{E}_r\)), and transmitted (\(\vec{E}_t\)) field in the graphene plane. Assuming that the contribution of the graphene susceptibility is a small perturbation of the total field, we set \(\mathcal{P}=0\) to calculate this symmetric surface field \(\vec{E}^{(s)}=\left[\vec{E}_i+\vec{E}_r+\vec{E}_t\right]/2\) in the axes of the sample (see Fig.~\ref{Fig:setup})
\begin{eqnarray}
 \begin{bmatrix}
  E^{(s)}_x\\E^{(s)}_y\\E^{(s)}_z
 \end{bmatrix}
 =\begin{bmatrix}
   \cos\alpha_t&0\\
   0&1\\
   -\sin\alpha_t(1+n_b^2/n_a^2)/2&0
  \end{bmatrix}
\cdot
 \begin{bmatrix}
 E^\parallel_t\\E^\perp_t
 \end{bmatrix},
\label{Eq:Es_of_Eparperp}
\end{eqnarray}
so that we can finally calculate the contribution of the graphene surface to the transmitted field, combining
\begin{equation}
 \mathcal{P}_i = \varepsilon_0\sum_{j=x,y,z} \chi_{ij}\otimes E^{(s)}_j,\;\left(i=x,y,z\right)
\label{Eq:P_of_Es}
\end{equation}
with (\ref{Eq:Es_of_Eparperp}), (\ref{Eq:transfer_par}), and (\ref{Eq:transfer_per}). 
In (\ref{Eq:P_of_Es}), \(\otimes\) denote the convolution product on time.
Note that due to~(\ref{Eq:Ffield}), \(\chi_{ij}\) is a surface quantity that could be linked to the volume quantity \(\chi^{v}_{ij}\) using \(\chi_{ij}=d\cdot\chi^{v}_{ij}\), where \(d\approx\mathrm{0.33~nm}\) is taken as the distance between two graphene sheets. For  \(\chi^{v}_{ij}\) in the order of unity, \(k_0\chi_{ij}\sim{}k_0d<10^{-3}\) in the visible and infrared regions.

This theory is valid for linear interactions. In the following section, we show that it can equally apply to the pump-probe geometry of Fig.~\ref{Fig:setup}, when we are interested in Kerr effect and two-photon absorption.

\subsection{Nonlinear interaction}

Most theoretical papers on graphene use the conductivity \(\sigma_{ij}\) rather than the susceptibility \(\chi_{ij}\). In what follows, we will consider the surface polarization that is linked to the surface current \(\vec{\mathcal{J}}=\partial_{t}\vec{\mathcal{P}}\), so that \(\sigma_{ij}(t)=\varepsilon_0\partial_t\chi_{ij}(t)\), or equivalently in the spectral domain \(\tilde{\sigma}_{ij}(\omega)=-i\varepsilon_0\omega\tilde{\chi}_{ij}(\omega)\).

To allow for a complete description of the material properties in the pump-probe geometry of Fig.~\ref{Fig:setup}, we model the third-order current density using 
\begin{equation}
 \mathcal{P}^{(3)}_i(t)
 =\varepsilon_0\sum_{\substack{j,k,l=\\x,y,z}}{\chi}^{(3)}_{ijkl}\stackrel{1}{\otimes}{E}^{(s)}_j
                                       \stackrel{2}{\otimes}{E}^{(s)}_k
                                       \stackrel{3}{\otimes}{E}^{(s)}_l,
 \label{Eq:NLresponse}
\end{equation}
where \(\stackrel{n}{\otimes}\) denotes the convolution on the \(n\)-th variable of \(\chi^{(3)}_{ijkl}(t_1,t_2,t_3)\),
and the electric-field components \(E^{(s)}_l\) can be written as the sum of paraxial pump (\(p\)) and probe (\(b\)) beams
\begin{equation}
E^{(s)}_l=E_l^{p,(s)}+E_l^{b,(s)}
         =\left[A_l^{p}\mathrm{e}^{\mathrm{i}\vec{k}_p\vec{r}}
               +A_l^{b}\mathrm{e}^{\mathrm{i}\vec{k}_b\vec{r}}\right] + \mathrm{(c.c.)},
\label{Eq:Efield}
\end{equation}
where \(\mathrm{(c.c.)}\) denotes the complex conjugate.

In the setup depicted on Fig.~\ref{Fig:setup}, the iris selects the output component collinear with the probe beam. Introducing~(\ref{Eq:Efield}) in (\ref{Eq:NLresponse}) shows that these components appear in terms containing one factor \(E_{\cdot}^b\), and either the pair \((E_{\cdot}^{b,(s)},E_{\cdot}^{b,(s)}{}^*)\) or \((E_{\cdot}^{p,(s)},E_{\cdot}^{p,(s)}{}^*)\), where \(\cdot\) denotes any index. Taking into account that the signal beam is very weak in comparison with the pump beam, we get
the spatially filtered contribution
\begin{equation} \mathcal{P}^{(3)}_{\vec{k}_s,i}(t)
 =6\, \varepsilon_0\!\sum_{\substack{j,k,l=\\x,y,z}}{\chi}^{(3)}_{ijkl}\stackrel{1}{\otimes}E^{b,(s)}_j
                                        \stackrel{2}{\otimes}{\left.E^{p,(s)}_k\right.^*
                                        \stackrel{3}{\otimes}E^{p,(s)}_l},
 \label{Eq:NLresponse_ks}
\end{equation}
in which the leading \(6\) comes from the intrinsic symmetries of the susceptibilities~\cite{Boyd201474}. From (\ref{Eq:NLresponse_ks}) we define a pseudo-linear susceptibility independent of the absolute phase of \(E^{p}\)
\begin{equation}
 \chi^{NL}_{ij}(t_1)=6\sum_{\substack{k,l=\\x,y,z}}{\chi}^{(3)}_{ijkl}(t_1,t_2,t_3)
                                       \stackrel{2}{\otimes}\left.E^{p,(s)}_k\right.^*
                                       \stackrel{3}{\otimes}E^{p,(s)}_l,
 \label{Eq:chiNLil}
\end{equation}
that depends on the pump signal shape and intensity, so that
\begin{equation}
 \mathcal{P}_{\vec{k}_s,i}=\varepsilon_0\hspace{-3mm}\sum_{j=x,y,z}\hspace{-2mm}\left[\chi_{ij}^{(1)}+\chi^{NL}_{ij}\right]\otimes{E^{b,(s)}_j}
 =\varepsilon_0\hspace{-3mm}\sum_{j=x,y,z}\hspace{-2mm}\chi_{ij}\otimes{E^{b,(s)}_j},
 \label{Eq:Jksi_chieff}
\end{equation}
which defines an effective first order susceptibility for \(E^{b,(s)}\).

\subsection{Optical Kerr effect}
Combining (\ref{Eq:P_of_Es}), (\ref{Eq:Es_of_Eparperp}), (\ref{Eq:transfer_par}), (\ref{Eq:transfer_per}), and (\ref{Eq:Jksi_chieff}), we find as transfer matrix between the incident and transmitted polarization components
\begin{equation}
 \begin{bmatrix}
  E^\parallel_t\\
  E^\perp_t
 \end{bmatrix}
=\mathcal{M}\cdot
 \begin{bmatrix}
  E^\parallel_i\\
  E^\perp_i
 \end{bmatrix}
=
 \begin{bmatrix}
  A&B\\
  C&D
 \end{bmatrix}
\cdot
 \begin{bmatrix}
  E^\parallel_i\\
  E^\perp_i
 \end{bmatrix}
,
\label{Eq:TransABCD}
\end{equation}
where
\begin{eqnarray}
{A}&=&P+PQ\chi_{x\parallel}+PR\chi_{z\parallel}
,\label{Eq:deltaA}\\
{B}&=&MQ\chi_{xy}+MR\chi_{zy}
,\label{Eq:deltaB}\\
{C}&=&PN\chi_{y\parallel}
,\label{Eq:deltaC}\\
{D}&=&M+MN\chi_{yy}
,\label{Eq:deltaD}\\
\chi_{i\parallel}&=&\cos\alpha_t\chi_{ix}-\sin\alpha_t(1+n_b^2/n_a^2)/2\,\chi_{iz}
.\label{Eq:chipar}
\end{eqnarray}
Measurement of the effects induced by the pump beam on the probe could be performed as follows. First, we switch the pump off which defines the transfer matrix \(\mathcal{M}_0\), and we set a normalized probe polarization state \(I_0=(E_i^\parallel,E_i^\perp)^t/E_i\),
that defines the normalized output state \(T_0=\mathcal{M}_0I_0/E_t\).
Then we tune the output analyzer to get zero transmitted signal. This is equivalent to project on the state \(T_\perp=(-T_0^\perp,T_0^\parallel)^\dagger\) orthogonal to \(T_0\). Finally, we switch the pump on, which induces a nonlinear change \(\delta\mathcal{M}=\mathcal{M}-\mathcal{M}_0\) on the transfer matrix, and we detect the power change at the output, given by
\begin{equation}
 \vert{E_{out}}\vert^2=\left\vert{}T_\perp^\dagger\delta\mathcal{M}I_0\right\vert^2\cdot\left\vert{}E^b_i \right\vert^2.
\end{equation}
This signal is very weak as it is proportional to \(\delta\mathcal{M}^2\sim{}(k_0d)^2<10^{-6}\), and it does not provide the sign of \(\delta\mathcal{M}\). To improve this, optical heterodyning is performed.

\subsection{Optical heterodyning}
We slightly change the input conditions on the quarter wave plate in Fig~\ref{Fig:setup}, or modify the analyzer angle, so that the input polarization state becomes
\(I_1=(I_0+i\theta{}I_\perp)/\sqrt{1+\vert\theta\vert^2}\), or the projection state becomes \(T_1=(T_\perp+\eta{}T_0)/\sqrt{1+\vert\eta\vert^2}\), where the terms proportional to \(\theta\) and \(\eta\) correspond to the local oscillator field~\cite{smith_optically-heterodyne-detected_2002,dremetsika_measuring_2016}. The measured output field is therefore
\begin{equation}
 \vert{E_{out}}\vert^2=\left\vert{}T_1^\dagger(\mathcal{M}_0+\delta\mathcal{M})I_1\right\vert^2
 \cdot\left\vert{}E^b_i \right\vert^2
.
\end{equation}
To isolate the weak nonlinear signal from the background, both the pump and probe beams are modulated at low frequencies \(\Omega_p\) and \(\Omega_b\) by means of a chopper~\cite{farrer_dynamics_1997}, and the signal power is measured, with a lock-in amplifier at \(\Omega_{bp}=\Omega_p+\Omega_b\), to the first order in \(\delta\mathcal{M}\). The 2D-OHD-OKE signal is
\begin{eqnarray}
S(\theta,\eta)&=&
2\mathrm{Re}\!\left[\left(T_1^\dagger\mathcal{M}_0I_1\right)^*
                                 \left(T_1^\dagger\delta\mathcal{M}I_1\right)
                           \right]\cdot\left\vert{}E^b_i\right\vert^2
.\label{Eq:SigOut}
\end{eqnarray}

In bulk OHD-OKE, \((\delta\!{\mathcal{M}})^2\) cannot be neglected as \(k_0d\geq1\), which leads to an additional ``homodyne'' signal.
The incident electric fields in medium (\(a\)) can be calculated from the incident power \(P\) and the effective beam area \(\pi{}w_0^2\) on the sample, using \(P^b=2\pi{}w_{0b}^2\varepsilon_0n_ac\vert{E^b_i}\vert^2\), and the same for \(P^p\).

\subsection{In-plane 2D-OHD-OKE}
\label{Sec:Classical_OKE}

As an example, we consider the in-plane 2D-OHD-OKE configuration depicted in Fig.~\ref{Fig:setup}, and used in Ref.~\onlinecite{dremetsika_measuring_2016}. The heterodyne parameters to measure the real part of the nonlinear response are \(\eta=0\), \(\theta\neq0\) , the refractive indices are \(n_a=1\) and \(n_b=1.5\) for a glass substrate. As the sample is not tilted, \(\alpha_i=0=\alpha_t\). We therefore have \(P=M=2/(1+n_b)\), \(N=Q=ik_0/(1+n_b)\), and \(R=0\). From (\ref{Eq:chipar}), we have \(\chi_{i\parallel}=\chi_{ix}\), and therefore
\begin{equation}
\mathcal{M}=2
 \begin{bmatrix}
  (1+n_b)+ik_0\chi_{xx} & \phantom{(1+n_b)}ik_0\chi_{xy}\\
  \phantom{(1+n_b)}ik_0\chi_{yx} & (1+n_b)+ik_0\chi_{yy}
 \end{bmatrix}/(1+n_b)^2.
\end{equation}
Pump and probe polarizations are depicted in the inset of Fig.~\ref{Fig:setup}. The input state is defined by \(I_0=(1,1)^t/\sqrt{2}\). When the pump is switched off, \(\chi_{ij}=\chi_{ij}^{(1)}=\chi_g\delta_{ij}\) due to the 6-th order symmetry of graphene, so that
\(\mathcal{M}_0\) is the identity matrix multiplied by \(2[1+ik_0\chi_g/(1+n_b)]/(1+n_b)\), 
which implies \(T_0=I_0\) and \(T_1=T_\perp=I_\perp=(-1,1)^{t}/\sqrt{2}\). 
As \(\mathcal{M}_0\) will be multiplied by a first order term in \(k_0d\), we can ignore the term proportional to \(k_0\chi_g\), so that \(T_1^\dagger\mathcal{M}_0I_1=
2i\theta/(1+n_b)\).

To evaluate \(\delta\mathcal{M}\), we calculate the pump field, which is  vertically polarized, orthogonal to its horizontal incidence plane. Therefore the symmetric surface field is given by \(E_t^{p,\perp}\) according to (\ref{Eq:Es_of_Eparperp}). As this beam makes an angle \(\alpha_i^p\) with \(\hat{z}\), using (\ref{Eq:transfer_per}) we get
 \(E_x^{p,(s)}=2E^{p}_i/(1+n_b\cos\alpha_t^p/\cos\alpha_i^p)\), with \(\alpha_t^p=\sin^{-1}(\sin\alpha_i^p/n_b)\). Using (\ref{Eq:chiNLil}) and the symmetries of the third-order susceptibility tensor, we get
\begin{eqnarray}
\delta\mathcal{M}&=&\frac{12ik_0}{(1+n_b)^2}
 \begin{bmatrix}
  \chi_{xxxx} & 0\\
  0 & \chi_{yyxx}
 \end{bmatrix}\left\vert{E_x^{p,(s)}}\right\vert^2,\\
 \mathrm{S}(\theta,0) &=& S_0
 \mathrm{Re}\!\left[
 \theta^*
 (\chi_{yyxx}-\chi_{xxxx})
 +i\vert{\theta}\vert^2
 (\chi_{xxxx}+\chi_{yyxx})
 \right]
\label{Eq:S_OKE_trad},\nonumber\\
S_0&=&\frac{96k_0\left\vert{E^{p}_iE^{b}_i}\right\vert^2}
       {(1+\vert\theta\vert^2)(1+n_b)^3(1+n_b\cos\alpha_t^p/\cos\alpha_i^p)^2}
,\label{Eq:S1}
\end{eqnarray}
where, for the sake of simplicity, we write \(\chi_{ijkl}=\chi^{(3)}_{ijkl}\).

Taking \(S(\theta,0)-S(-\theta,0)\) we get access to
\( \mathrm{Re}\!\left[\chi_{xxxx}-\chi_{yyxx} \right]= \mathrm{Re}\!\left[\chi_{xyxy}+\chi_{xyyx} \right]\), where we have used symmetry relations. 

Using (\ref{Eq:S1}) and the 
data from Ref.~\onlinecite{dremetsika_measuring_2016}, we calculate
\(\mathrm{Re}\left[\chi_{xyxy}+\chi_{xyyx}\right] = -200\,\mathrm{nm^3/V^2}\).

\section{Experimental results}

In our experiments we used a monolayer graphene film on glass from Graphene Laboratories Inc., which was grown by catalyzed chemical vapor deposition (CVD). We verified that the glass substrate does not present a nonlinear response.

Our experimental setup is based on the one depicted in Ref.~\onlinecite{dremetsika_measuring_2016}, with the modifications appearing on Fig.~\ref{Fig:setup}.
The 180-fs pulses at 1600\,nm derived from an optical parametric oscillator (OPO), pumped by a Ti:Sapphire laser, with a repetition rate of  82\,MHz. The pump-probe power ratio is tuned around 15:1. The pump and probe beams are spatially overlapped on the sample and focused down to a beam waist of \(w_p\approx20\,\mbox{\textmu{m}}\) and \(w_b\approx15\,\mbox{\textmu{m}}\) respectively. The pump intensity is set around \(5\cdot10^{12}\,\mathrm{W/m^2}\), which is far below the damage threshold of graphene~\cite{currie_quantifying_2011}. The effective interaction length was \(L\approx100\,\mbox{\textmu{m}}\). 
The heterodyne parameters were either \(\theta=\pm\tan4\mbox{\textdegree}\), \(\eta=0\),  or \(\eta=\pm\tan4\mbox{\textdegree}\), \(\theta=0\). 
The angle between the pump and the probe beam is \(34\)\textdegree. For the out-of-plane measurements, the sample was rotated with a precision rotation mount.
As in Ref.~\onlinecite{dremetsika_measuring_2016}, we used a Silicon reference sample. 

\subsection{Temporal response of the tensor components}
\begin{figure}[h]
\centering
\begin{overpic}[width=0.49\linewidth]{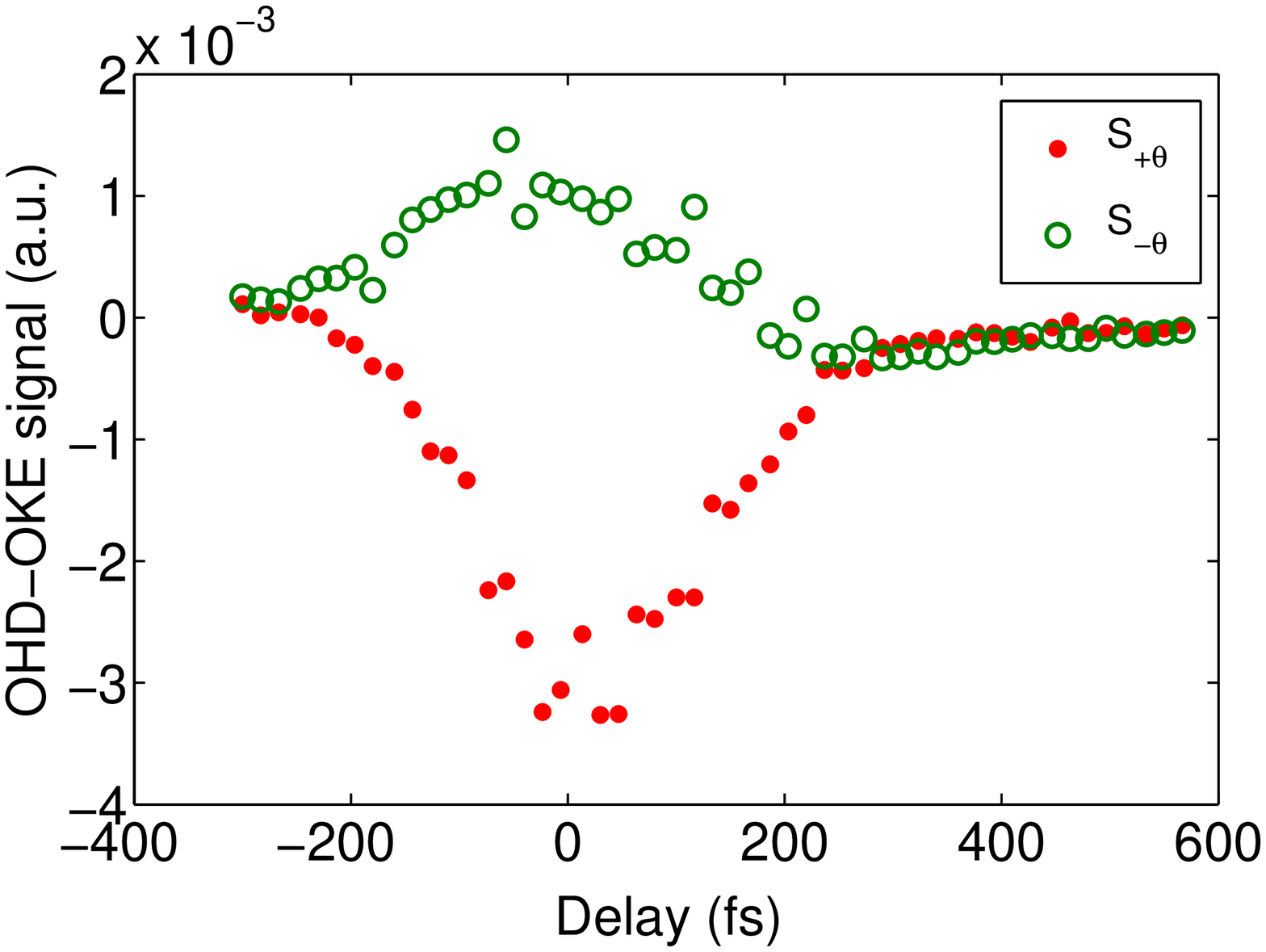}
\put(15,60){(a)}
\end{overpic}
\hfill
\begin{overpic}[width=0.49\linewidth]{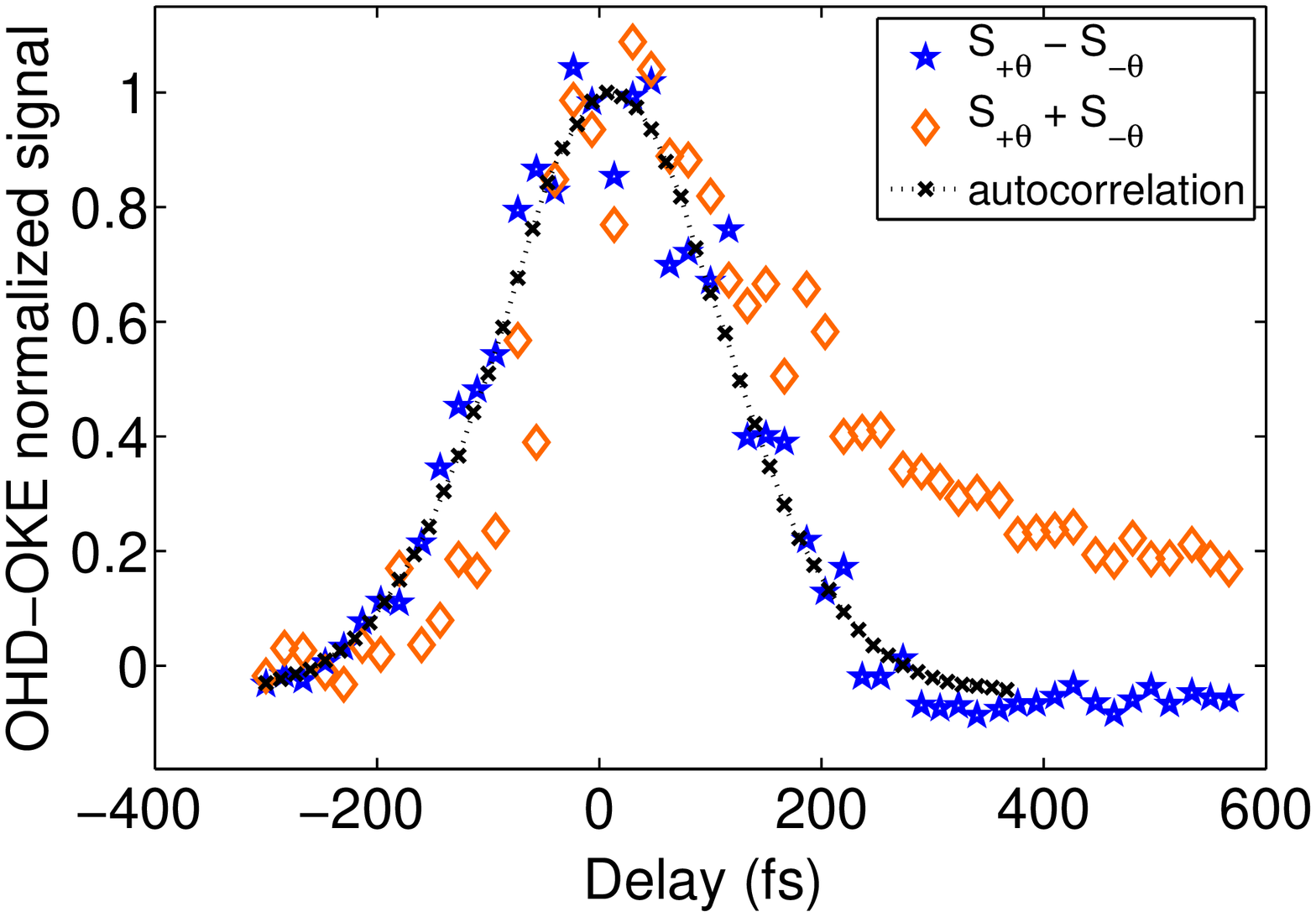}
\put(15,60){(b)}
\end{overpic}

\caption{Experimental results: (a)~2D-OHD-OKE signal of graphene at positive and negative heterodyne angle. (b)~Normalized difference and sum of 2D-OHD-OKE signals from (a)~compared with the pulse autocorrelation obtained by the OHD-OKE signal from the Silicon reference sample.}
\label{fig:OKEresults}
\end{figure}

Starting with the configuration described in Sec.~\ref{Sec:Classical_OKE}, we recorded the temporal response of the 2D-OHD-OKE signal shown in Fig.~\ref{fig:OKEresults}(a). At long delay, both signals are equal, which is explained by studying the different tensor components, namely by taking the difference and the sum of the signals \(S(\theta,0)\), and \(S(-\theta,0)\), from~(\ref{Eq:S1}). As shown on Fig.~\ref{fig:OKEresults}(b),  the first signal, \(S^\theta_{dif} \propto  \theta \mathrm{Re}\!\left[\chi_{xyxy}+\chi_{xyyx}\right]\), which is purely due to induced birefringence, has a fast response. This is indeed demonstrated by the perfect fit with the autocorrelation trace of the input pulses, that implies a relaxation time shorter than the 180-fs pulse duration. The second signal, \(S^\theta_{sum} \propto  \theta^2 \,\mathrm{Im}\!\left[\chi_{xyxy}+\chi_{xyyx}+2\chi_{xxyy}\right]\), appears due to the nonlinear absorption of the local oscillator field and therefore is negligible for materials with weak nonlinear absorption. Here, this signal is important, and presents a completely different behavior from the birefringent signal, with a picosecond relaxation time, characteristic of graphene~\cite{breusing_ultrafast_2011}.

\begin{figure}
\centering
\begin{overpic}[width=0.49\linewidth]{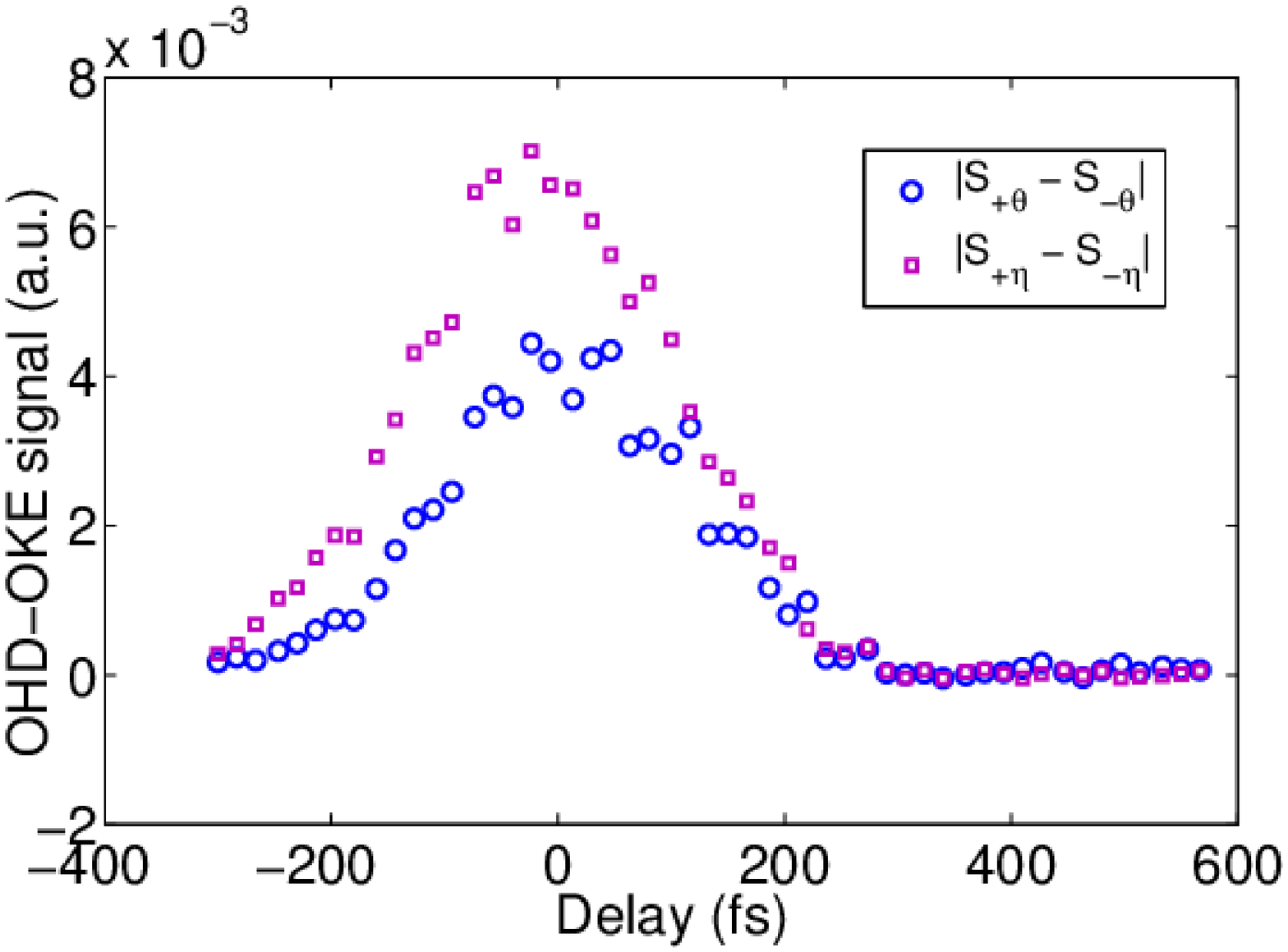}
\put(15,60){(a)}
\end{overpic}
\hfill
\begin{overpic}[width=0.49\linewidth]{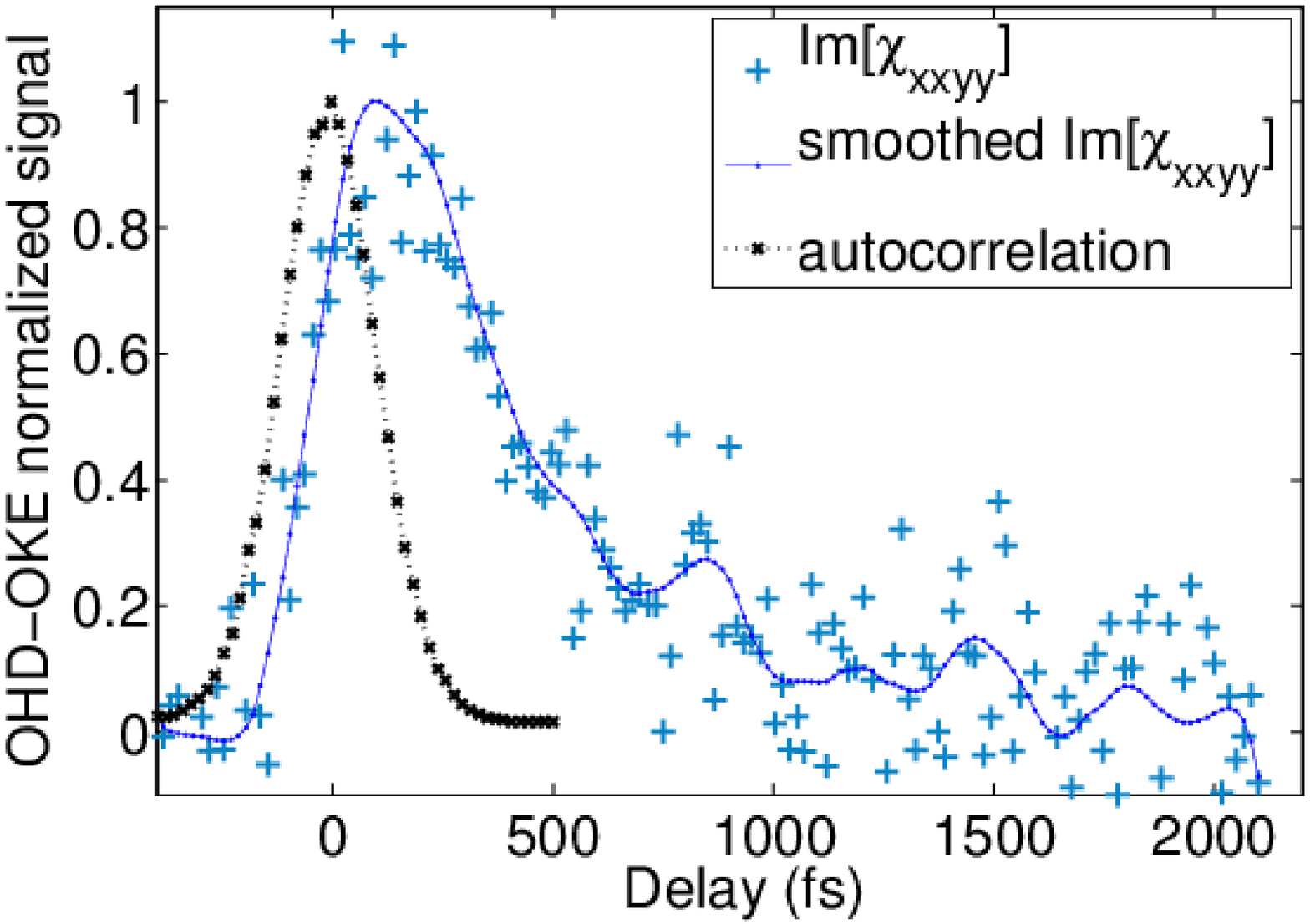}
\put(14,60){(b)}
\end{overpic}
\caption{Experimental results: (a)~comparison between \(\vert{S_{dif}^{\theta}}\vert\) and \(\vert{S_{dif}^{\eta}}\vert\); 
(b)~normalized \(S_{sum}^{\theta}\) for parallel pump and probe polarizations, providing \(\mathrm{Im}[\chi_{xxyy}]\) compared with the autocorrelation. Smoothed data (Savitzky-Golay) are provided as guide for the eye.}
\label{fig:OKEresults2}
\end{figure}

Next, we studied induced dichroism, which corresponds to the imaginary part of the nonlinearity, similarly to the birefringence by considering \(S(0,\eta)\) and~\(S(0,-\eta)\) (see Fig.~\ref{fig:OKEresults2}(a)). The induced dichroism \(S^\eta_{dif} \propto  \eta \,\mathrm{Im}\!\left[\chi_{xyxy}+\chi_{xyyx}\right]\) is also characterized by a fast relaxation, which is in agreement with the conclusion of Mittendorf \textit{et al.}~\cite{mittendorff_anisotropy_2014}, that the anisotropic distribution of photo-excited carriers in graphene has a fast relaxation time of 150\,fs. This anisotropic distribution is actually the microscopic origin of the induced dichroism or birefringence. 
The expression (and the observed behavior) of \(S^\eta_{sum}\) is the same as \(S^\theta_{sum}\). 
The ratio from the nonlinear dichroic losses to the nonlinear birefringence is evaluated to \(S^\eta_{dif}/S^\theta_{dif}\approx1.6\), which provides \(\mathrm{Im}\left[\chi_{xyxy}+\chi_{xyyx}\right]\approx-320\,\mathrm{nm^3/V^2}\), at zero pump-probe delay.

We compared the in-plane OHD-OKE data of graphene and Silicon, as we did in Ref.~\onlinecite{dremetsika_measuring_2016}, for the real part of the nonlinearity. We verified that the two-photon absorption coefficient of Silicon 
is in agreement with published values. As for the refractive part~\cite{dremetsika_measuring_2016}, the signals of graphene and Silicon presented opposite signs, which is not surprising, since it is well known that graphene is a saturable absorber.
Saturable absorption does not scale linearly with the input intensity, so the measured \(\mathrm{Im}\!\left[\chi_{xyxy}+\chi_{xyyx}\right]\) should decrease with increasing intensity. This was experimentally confirmed.

Finally, the imaginary part of \(\chi_{xxyy}\) was studied separately by taking measurements in a different configuration, in which the input pump and probe polarizations are parallel. In this case,  we get \(S^\theta_{sum}\propto  \theta^2 \,\mathrm{Im}\!\left[\chi_{xxyy}\right]\), which is shown in Fig.~\ref{fig:OKEresults2}(b). We infer that the relaxation time of \(\mathrm{Im}\!\left[\chi_{xxyy}\right]\) is around 1\,ps. By comparing the magnitude of the signals, we find \(\mathrm{Im}\!\left[\chi_{xxyy}\right] \approx 1.7\, \mathrm{Im}\!\left[\chi_{xyxy}+\chi_{xyyx}\right]\). We note that \(\mathrm{Re}\!\left[\chi_{xxyy}\right]\) is not accessible with simple experimental configurations, but it has probably the same temporal dependence as \(\mathrm{Im}\!\left[\chi_{xxyy}\right]\). At zero pump-probe delay, we find \(\mathrm{Im}\!\left[\chi_{xxyy}\right] \approx-540\,\mathrm{nm^3/V^2}\).

\subsection{Out-of-plane tensor components}
\begin{figure}
 \begin{overpic}[width=0.49\linewidth]{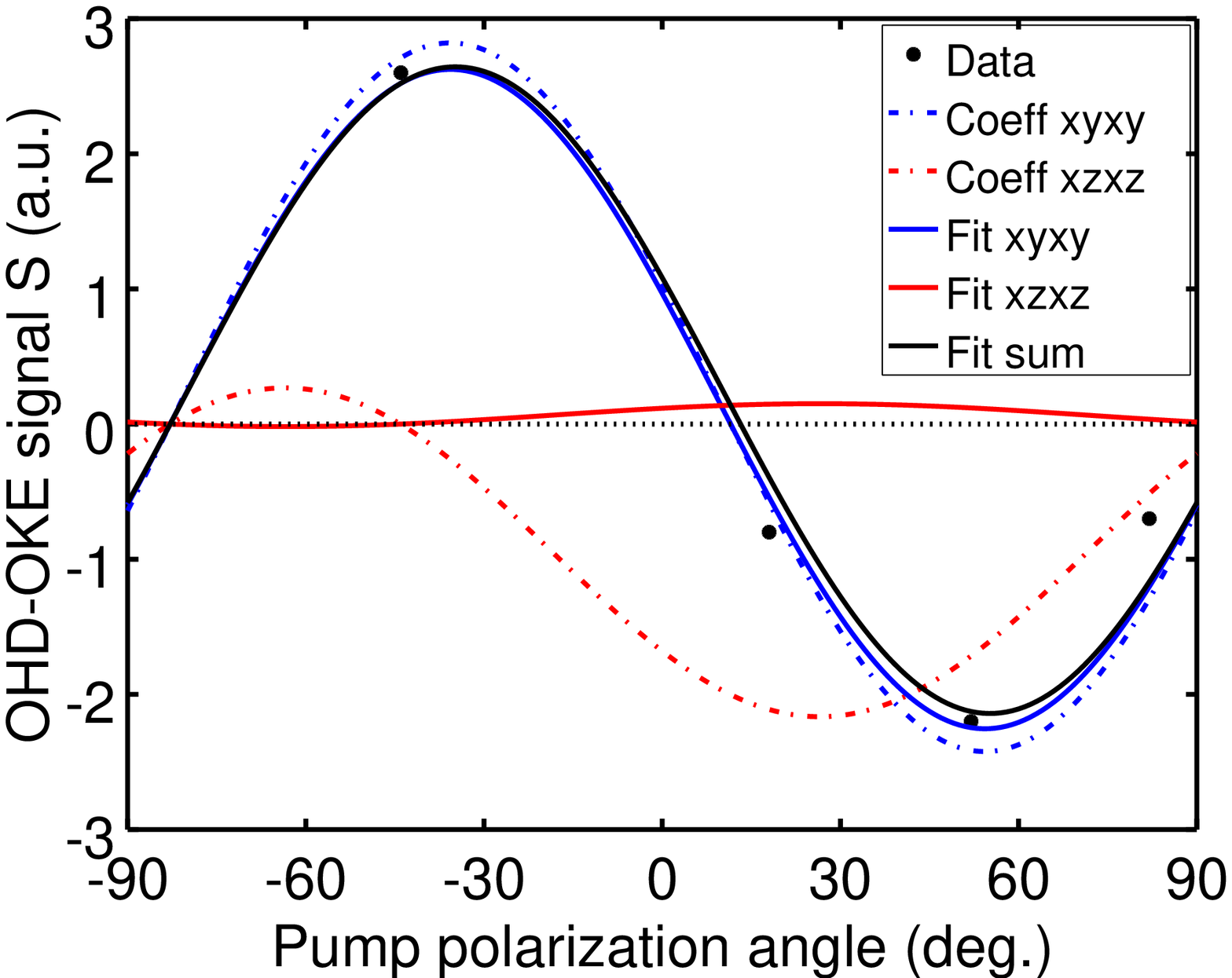}
 \put(15,60){(a)}
 \end{overpic}\hfill
 \begin{overpic}[width=0.49\linewidth]{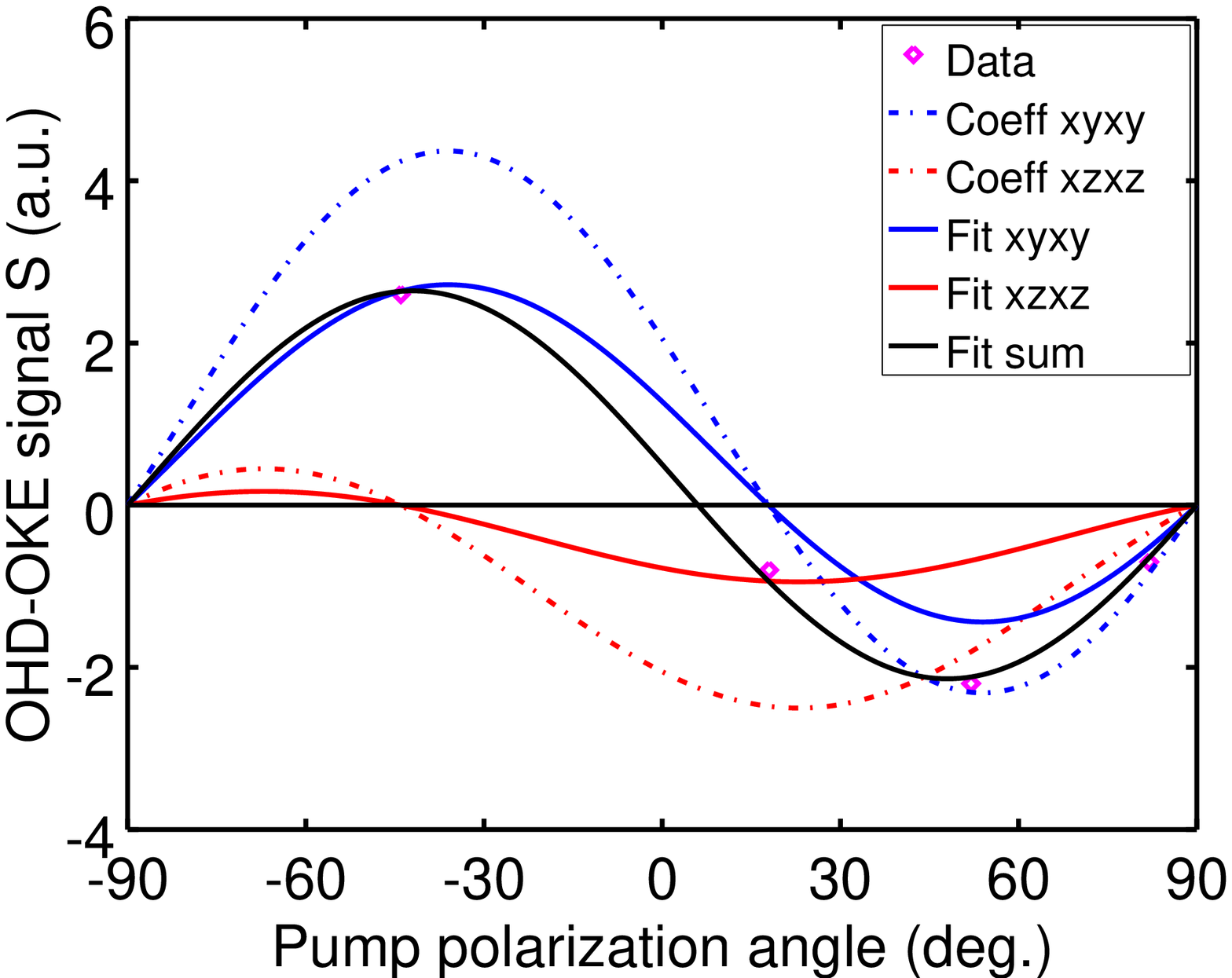}
 \put(15,60){(b)}
 \end{overpic}
\caption{Experimental data and fitting of the odd part of the 2D-OHD-OKE signal for a vertical input polarization of the probe (\(x\)-axis). The linear pump polarization is continuously set  from vertical (\(-90\)\textdegree) to horizontal (\(0\)\textdegree) and vertical (\(90\)\textdegree). \(C_{xyxy}\) and \(C_{xzxz}\) denote respectively the coefficients of the real parts of \(\chi_{xyxy}+\chi_{xyyx}\), and \(\chi_{xzxz}+\chi_{xzzx}\). Data are fitted to these curves: (black) total fit, (blue) xyxy component, (red) xzxz component. Coefficients \(C_{xyxy}\) and \(C_{xzxz}\) are calculated using (a) the symmetric surface field; (b) without taking the field in the substrate into account.}
\label{Fig:outofplane}
\end{figure}

To measure the out-of-plane tensor components, we tilt the sample around the horizontal axis, so that \(\alpha_i=30\)\textdegree, and we calculate the coefficients \(C_{ijkl}\) appearing in \(S(\theta,0)-S(-\theta,0)=S_1\mathrm{Re}\!\left[2\theta^*\sum_{ijkl=x,y,z}C_{ijkl}\chi_{ijkl}\right]\). 
When the input polarization of the probe beam is set vertical or horizontal, only four coefficients appear in \(S(\theta,0)-S(-\theta,0)\): namely \(C_{xyxy}=C_{xyyx}\), and \(C_{xzxz}=C_{xzzx}\) or \(C_{zxzx}=C_{zxxz}\) respectively. With the probe polarized vertically their variation with the polarization of the pump beam is shown in Fig.~\ref{Fig:outofplane} (dotted-dashed lines). Experimental data appear as black dots. In panel (a), the symmetric surface field is used, as explained in Sec.~\ref{Sec:LinearMaterialResponse}, while in panel (b), the expressions for bulk materials are used.
The relative values of  \(\mathrm{Re}\!\left[\chi_{xyxy}+\chi_{xyyx}\right]\) and \(\mathrm{Re}\!\left[\chi_{xzxz}+\chi_{xzzx}\right]\) are obtained by fitting the experimental data. In Fig.~\ref{Fig:outofplane}, the black curve shows the fitted curve, while the blue and the red ones show the contributions of the two sets of susceptibilities to the total curve. In panel (a) we see that the contribution \(xzxz\) to the fit is very weak, and is not needed to explain the experimental data, as its amplitude is in the order of the experimental error. Setting \(\eta=\pm\tan4\)\textdegree\ and \(\theta=0\), \(S(0,\eta)-S(0,-\eta)\) provides the ratio of the imaginary parts of the same components, which demonstrates that
\(\left\vert\chi_{xzxz}+\chi_{xzzx}\right\vert/\left\vert\chi_{xyxy}+\chi_{xyyx}\right\vert<0.1.\)
Using a second set of experimental data, with the probe beam polarized horizontally, we reach similar conclusions for the real and imaginary parts of \(\chi_{zxzx}+\chi_{zxxz}\). These results validate the theoretical assumptions that \(\chi_{xzxz}\), \(\chi_{xzzx}\),\(\chi_{zxzx}\), and \(\chi_{zxxz}\), are negligible~\cite{cheng_third-order_2015}, with a magnitude that does not exceed \(20\,\mathrm{nm^3/V^2}\).
Fig.~\ref{Fig:outofplane}(b) shows that the use of a bulk theory neglecting the substrate,
provides a ratio around 0.5, which would lead to an opposite conclusion.

\section{Discussion}

\subsection{In-plane components}
In order to compare our values of the in-plane components with other values from the literature that are mostly reported as volume susceptibilities or as an effective nonlinear refractive index, we provide the appropriate conversions in Tab.~\ref{Tab}.
\begin{table}[h]
\caption{Estimated parameters of the third-order optical nonlinearity from 2D-OHD-OKE. The real part of the complex nonlinear index \(n_{2c}\) corresponds to the effective nonlinear refractive index measured in a previous work~\cite{dremetsika_measuring_2016}. Conversion between this value and the surface susceptibility is performed using \(k_0 d n_2 P^p\) for the phase shift. All the parameters correspond to the surface components \(\chi_{xyxy}+\chi_{xyyx}\). They are obtained with a pump intensity around \(5\cdot10^{12}\,\mathrm{W/m^2}\). The real (resp. imaginary) part of \(\sigma^{(3)}\) is calculated from the imaginary (resp. real) part of \(\chi^{(3)}\).} 
\label{Tab}
\begin{center}
\begin{tabular}{c*{5}{|c}}
      &\(n_{2c}\)  & \(\chi^{(3)}_s\)& \(\chi^{(3)}_v\)& \(\chi^{(3)}_v\)&\(\sigma^{(3)}\)\\
\hline
 units &\textmu\(\mathrm{m^2/W}\)   & \(\mathrm{nm^3/V^2}\)&  \(\mathrm{nm^2/V^2}\)&\( \) esu &\(\mathrm{Am^2/V^3}\)\\
\hline
 Real      &-0.1 & -200 & -600 & -\( 0.4\cdot10^{-7}\) &-\(3.3\cdot10^{-21}\)\\
 Imag.     &-0.16& -320 & -960 & -\( 0.65\cdot10^{-7}\) &\(2.1\cdot10^{-21}\)
\end{tabular}
\end{center}
\end{table}%
  
Comparison of the effective nonlinear refractive index with the earlier values from Z-scan~\cite{zhang_z-scan_2012,chen_nonlinear_2013} was provided in a previous work~\cite{dremetsika_measuring_2016}. More recent reported values~\cite{demetriou_nonlinear_2016,PhysRevApplied.6.044006} are in good agreement, in sign and magnitude, with those already published values. It should be noted that Z-scan probes the nonlinearity related to \(\mathrm{Re}\left[\chi_{xxxx}\right]\), so that we do not expect an exact match of the results. This is also true for results from four-wave mixing (FWM) experiments.
Because we have not measured \(\mathrm{Re}\left[\chi_{xxyy}\right]\), we estimate the magnitude of the volume susceptibility \(\vert\chi_{v}^{(3)}\vert\), either by neglecting the \(\chi_{xxyy}\) component, or by assuming that \(\mathrm{Re}\left[\chi_{xxyy}\right]\)  is on the order of \(\mathrm{Re}\!\left[\chi_{xyxy}+\chi_{xyyx}\right]\). In both cases, we get an absolute value in the order of \(10^{-7}\mathrm{esu}\), which is in agreement with the value from Ref.\,\onlinecite{hendry_coherent_2010}.

To compare our results with theoretical values, we need to estimate the doping level of our graphene sample. Graphene deposited on glass or silicon substrates is low p-doped, as shown, for example, in Ref.\,\onlinecite{alexander_electrically_2017}, where the chemical potential is evaluated between -0.3\,eV and -0.2\,eV. At those doping levels and the wavelength used in our experiments (1600\,nm, or \(\hbar\omega\approx0.8\,\)eV) the nonlinear conductivity in Ref.~\onlinecite{cheng_third_2014} diverges. Assuming higher doping levels, so that \(\hbar\omega/\vert\mu\vert< 2\), Cheng and co-workers estimated a nonlinear refractive index two orders of magnitude lower than our value. 
In a later theoretical work~\cite{cheng_third-order_2015}, the same authors added phenomenological relaxation parameters and finite temperature in their theory. To compare our values with these more recent theoretical predictions, we refer to Tab.~\ref{Tab}.
The predictions for chemical potential \(\vert\mu\vert=\)0.3\,eV, presented in Fig. 4 of Ref.\,\onlinecite{cheng_third-order_2015}, show higher values of Re\([\sigma^{(3)}]\) than ours, with a discrepancy varying from one to two orders of magnitude depending on the phenomenological relaxation parameter introduced in the theory. Predictions from Ref.\,\onlinecite{mikhailov_quantum_2016} for the same parameters agree with those of Cheng and coworkers~\cite{cheng_third-order_2015}.  
Finally, in Ref.\,\onlinecite{alexander_electrically_2017}, theoretical values are compared to measurements performed with FWM on chip in continuous regime. The theoretical values of the third-order surface conductivity are estimated around \(10^{-18}\mathrm{Am^2/V^3}\), three orders of magnitude higher than our values (Tab.~\ref{Tab}), while the experimental values are around \(10^{-19}\mathrm{Am^2/V^3}\), in better agreement with our results. The high discrepancy between the experimental results can be due to many factors, such as the continuous regime and the waveguiding geometry. 

To conclude this discussion, the values that we measure match the values reported in other experimental works clearly better, than those reported in theoretical studies. Possible explanations of these differences between theory and experiments are discussed in Ref.\,\onlinecite{cheng_third-order_2015}.

\subsection{Out-of-plane components}
To verify the theoretical assumption that the out-of-plane components of the susceptibilities can be neglected~\cite{cheng_third-order_2015}, we should either consider to work with sheet currents, with the problem that textbook expressions do not allow to calculate the impact of the out-of-plane components, and limit therefore the possibility to measure them; or we could use a 3D theory that neglects the local field and relates microscopic parameters to light propagation in an homogeneous medium. We have shown how to circumvent this problem to measure the out-of-plane components, and show that \(\chi_{xzxz}+\chi_{xzzx}\), and \(\chi_{zxzz}+\chi_{zxxz}\) are negligible.

\subsection{Temporal response}

Measuring \(\chi_{xxxx}\) is possible with one-beam techniques involving a single polarization. However, it does not offer the possibility to separate different contributions with different time responses. The 2D-OHD-OKE method allows to record the temporal response of \(\chi_{xyxy}+\chi_{xyyx}\), and \(\chi_{xxyy}\) separately.

\subsection{3D versus 2D material parameters}
Nonlinear optical properties are usually measured through amplitude changes, that can find their origin in phase changes. Going from the amplitude and phase variations to the tensor components requires the use of a model. In this paper, we provide a complete analytical model from the Maxwell's equation to the measured power (\ref{Eq:SigOut}). This expression differs from the one used in a bulk material.
Therefore, we should refrain from using equations relating experimental phase changes to 3D propagation parameters in order to estimate the susceptibilities. Doing so would, for example, introduce the refractive index of the 2D material under consideration, while the expressions based on the current sheets would not. Indeed, our results show the possibility to measure the third-order surface susceptibilities (or the associated conductivities) considered in theoretical works~\cite{cheng_third_2014,mikhailov_quantum_2016}, without estimating the refractive index of the 2D material.

\section{Conclusion} 

In order to verify that the out-of-plane components of the third-order nonlinear optical susceptibility of graphene are negligible, we have developed the 2D-OHD-OKE method, and an appropriate model for the optical interaction of light at an interface with a 2D material. Six new values for the real and imaginary parts of these components for graphene  have been provided at zero pump-probe delay, together with their time evolution. The out-of plane components that we measured are negligible.

We have shown that \(\chi_{xyxy}+\chi_{xyyx}\) accounts for the fast birefringent and dichroic contribution to the nonlinear response, in agreement with~\cite{mittendorff_anisotropy_2014} for the dichroic response. We have compared its magnitude with \(\chi_{xxyy}\) which has a slower (ps) relaxation time.

Equation (\ref{Eq:SigOut}) allows to calculate the intrinsic parameters from the experimental data of 2D-OHD-OKE. It reveals the  importance to take the substrate into account, as it modifies the symmetric surface field. Our modeling can be used to discriminate the optical interactions between a 2D material and its substrate from the chemical ones. It could also help for the development of numerical simulation tools using sheet currents to model the nonlinear optical response of graphene.

The 2D-OHD-OKE method presented here should apply to all 2D materials, and provide an efficient mean to retrieve experimentally their fundamental parameters.

\begin{acknowledgments}
This work is partially supported by the  Belgian Science Policy Office (BELSPO) under grant IAP7-35. E. D. is funded by the Fund for Research Training in Industry and Agriculture (FRIA, Belgium).   
\end{acknowledgments}

\end{document}